\documentclass[10pt]{IEEEtran}

\usepackage{amsmath,amssymb,graphicx,cite,bm}

\renewcommand{\(}{\left(}
\renewcommand{\)}{\right)}
\renewcommand{\[}{\left[}
\renewcommand{\]}{\right]}
\renewcommand{\c}{\mathbf{c}}
\renewcommand{\a}{\mathbf{a}}
\renewcommand{\b}{\mathbf{b}}

\newcommand{\Sig}{\bm{\Sigma}}

\newcommand{\y}{\mathbf{y}}

\newcommand{\0}{\mathbf{0}}

\newcommand{\x}{\mathbf{x}}
\newcommand{\I}{\mathbf{I}}

\newcommand{\X}{\mathbf{X}}
\newcommand{\Y}{\mathbf{Y}}

\renewcommand{\b}{\mathbf{b}}

\newtheorem{lemma}{Lemma}

\title{A greedy approach to sparse canonical correlation analysis}

\author{Ami Wiesel, Mark Kliger and Alfred O. Hero III\\
Dept. of Electrical Engineering and Computer Science\\
University of Michigan, Ann Arbor, MI 48109-2122, USA\\
\thanks{The first two authors contributed equally to this manuscript.
This work was supported in part by an AFOSR MURI under Grant
FA9550-06-1-0324.}}

\begin{document}
 \maketitle

\begin{abstract}
  We consider the problem of sparse canonical correlation analysis (CCA),
  i.e., the search for two linear combinations, one for each
  multivariate, that yield maximum correlation using a specified number of variables.
  We propose an efficient numerical approximation based on a direct greedy approach which bounds the correlation at each stage. The method is specifically designed to cope with large data sets and its computational complexity depends only on the sparsity levels. We analyze
  the algorithm's performance through the tradeoff between correlation and parsimony. The results of numerical simulation
  suggest that a significant portion of the correlation may be captured
  using a relatively small number of variables. In addition, we examine the use of sparse CCA as a
  regularization method when the number of available samples is small compared to the dimensions of the multivariates.
\end{abstract}

\section{Introduction}

Canonical correlation analysis (CCA), introduced by Harold Hotelling
\cite{Hotelling36}, is a standard technique in multivariate data
analysis for extracting common features from a pair of data sources
\cite{thompson:84,Anderson2003}. Each of these data sources generates a random
vector that we call a multivariate. Unlike classical dimensionality
reduction methods which address one multivariate, CCA takes into
account the statistical relations between samples from two
spaces of possibly different dimensions and structure. In
particular, it searches for two linear combinations, one for each
multivariate, in order to maximize their correlation. It is used in
different disciplines as a stand-alone tool or as a preprocessing
step for other statistical methods. Furthermore, CCA is a generalized framework which
includes numerous classical methods in statistics, e.g., Principal
Component Analysis (PCA), Partial Least Squares (PLS) and Multiple
Linear Regression (MLR) \cite{borga-unified}.  CCA has recently regained attention with the advent of
kernel CCA and its application to independent component analysis
\cite{bach:2003,Gretton:2005}.

The last decade has witnessed a growing interest in the search for
sparse representations of signals and sparse numerical methods.
Thus, we consider the problem of sparse CCA, i.e., the search for
linear combinations with maximal correlation using a small number of
variables. The quest for sparsity can be motivated through
various reasonings. First is the ability to interpret
and visualize the results.  A small number of variables
allows us to get the ``big picture'', while sacrificing some of the
small details. Moreover, sparse representations enable the use of
computationally efficient numerical methods, compression techniques,
as well as noise reduction algorithms. The second motivation for
sparsity is regularization and stability. One of the main
vulnerabilities of CCA is its sensitivity to a small number of
observations. Thus, regularized methods such as ridge
CCA \cite{vinod:76} must be used. In this context, sparse CCA is
a subset selection scheme which allows us to reduce the
dimensions of the vectors and obtain a stable
solution.

To the best of our knowledge the first reference to sparse CCA
appeared in \cite{thompson:84} where backward and stepwise subset
selection were proposed. This discussion was of qualitative nature
and no specific numerical algorithm was proposed. Recently,
increasing demands for multidimensional data processing and
decreasing computational cost has caused the topic to rise to
prominence once again
\cite{hardoontr07,aspremontgjl05,waaijenborg:2007,Parkhomenko:2007,FyfeL06,TanF01}.
The main disadvantages with these current solutions is that there is
no direct control over the sparsity and it is difficult (and
non-intuitive) to select their optimal hyperparameters. In addition,
the computational complexity of most of these methods is too high
for practical applications with high dimensional data sets. Sparse
CCA has also been implicitly addressed in
\cite{Sriperumbudur:07,aspremontgjl05} and is intimately related to
the recent results on sparse PCA
\cite{Hastie_SPCA,moghadan:2006,aspremontgjl05,daspremont-2007}.
Indeed, our proposed solution is an extension of the results in
\cite{daspremont-2007} to CCA.

The main contribution of this work is twofold. First, we derive CCA
algorithms with direct control over the sparsity in each of the multivariates and examine their
performance. Our computationally efficient methods are specifically
aimed at understanding the relations between two data sets of large
dimensions. We adopt a forward (or backward) greedy approach which
is based on sequentially picking (or dropping) variables. At each
stage, we bound the optimal CCA solution and bypass the need to
resolve the full problem. Moreover, the computational complexity of
the forward greedy method does not depend on the dimensions of the
data but only on the sparsity parameters. Numerical simulation
results show that a significant portion of the correlation can be
efficiently captured using a relatively low number of non-zero
coefficients. Our second contribution is investigation of sparse CCA
as a regularization method. Using empirical simulations we examine
the use of the different algorithms when the dimensions of the
multivariates are larger than (or of the same order of) the number
of samples and demonstrate the advantage of sparse CCA. In this
context, one of the advantages of the greedy approach is that it
generates the full sparsity path in a single run and allows for efficient parameter
tuning using cross validation.

The paper is organized as follows. We begin by describing the
standard CCA formulation and solution in Section \ref{sec_review}.
Sparse CCA is addressed in Section \ref{sec_sparse} where we review
the existing approaches and derive the proposed greedy method. In
Section \ref{sec_num}, we provide performance analysis using
numerical simulations and assess the tradeoff between correlation
and parsimony, as well as its use in regularization. Finally,
a discussion is provided in Section \ref{sec_conc}.

The following notation is used. Boldface upper case letters denote
matrices, boldface lower case letters denote column vectors, and
standard lower case letters denote scalars. The superscripts
$(\cdot)^T$ and $(\cdot)^{-1}$ denote the transpose and inverse operators, respectively.
By $\I$ we denote the identity matrix. The operator $\|\cdot\|_2$ denotes the L2 norm, and $\|\cdot\|_0$ denotes the cardinality operator. For two sets of indices $I$ and $J$, the matrix $\X^{I,J}$ denotes
the submatrix of $\X$ with the rows indexed by $I$ and columns
indexed by $J$. Finally, $\X\succ\0$ or $\X\succeq\0$ means that the matrix $\X$ is positive definite or positive semidefinite, respectively.

\section{Review on CCA}\label{sec_review}
In this section, we provide a review of classical CCA. Let $\x$ and $\y$ be two zero mean random vectors of lengths $n$ and
$m$, respectively, with joint covariance matrix:
\begin{eqnarray}
  \Sig=\[\begin{array}{ll}
           \Sig_x & \Sig_{xy} \\
           \Sig_{xy}^T & \Sig_y
         \end{array}
  \]\succeq\0,
\end{eqnarray}
where $\Sig_x$, $\Sig_y$ and $\Sig_{xy}$ are the covariance of $\x$,
the covariance of $\y$, and their cross covariance, respectively.
CCA considers the problem of finding two linear
combinations $X=\a^T\x$ and $Y=\b^T\y$ with maximal
correlation defined as
\begin{eqnarray}
\frac{\text{cov}\{X,Y\}}{\sqrt{\text{var}\{X\}}\sqrt{\text{var}\{Y\}}},
\end{eqnarray}
where $\text{var}\{\cdot\}$ and $\text{cov}\{\cdot\}$ are the variance and covariance operators, respectively, and we define $0/0=1$. In terms of $\a$ and $\b$ the correlation can be easily expressed as
\begin{eqnarray}
\frac{\a^T\Sig_{xy}\b}{\sqrt{\a^T\Sig_x\a}\sqrt{\b^T\Sig_y\b}}.
\end{eqnarray}
Thus, CCA considers the following optimization problem
\begin{eqnarray}\label{cca}
 \rho\(\Sig_x,\Sig_y,\Sig_{xy}\)=\max_{\a\neq\0,\b\neq\0}\frac{\a^T\Sig_{xy}\b}{\sqrt{\a^T\Sig_x\a}\sqrt{\b^T\Sig_y\b}}.
\end{eqnarray}
Problem (\ref{cca}) is a multidimensional non-concave maximization
and therefore appears difficult on first sight. However, it has a
simple closed form solution via the generalized eigenvalue
decomposition (GEVD). Indeed, if $\Sig\succ\0$, it is easy to show that the optimal $\a$ and $\b$ must satisfy:
\begin{eqnarray}\label{gev}
 \[\begin{array}{cc}
     \0 & \Sig_{xy} \\
     \Sig_{xy}^T & \0
   \end{array}
 \]\[\begin{array}{c}
       \a \\
       \b
     \end{array}
 \]=\lambda\[\begin{array}{cc}
     \Sig_{x} & \0\\
     \0 & \Sig_{y}
   \end{array}
 \]\[\begin{array}{c}
       \a \\
       \b
     \end{array}
 \],
\end{eqnarray}
for some $0\leq\lambda\leq 1$. Thus, the optimal value of
(\ref{cca}) is just the  principal generalized eigenvalue
$\lambda_{\max}$ of the pencil (\ref{gev}) and the optimal solution
$\a$ and $\b$ can be obtained by appropriately partitioning the
associated eigenvector. These solutions are invariant to scaling of
$\a$ and $\b$ and it is customary to normalize them such that
$\a^T\Sig_x\a=1$ and $\b^T\Sig_y\b=1$.
%\begin{eqnarray}
%  \text{var}\{X\}&=&\a^T\Sig_x\a=1\nonumber\\
%  \text{var}\{Y\}&=&\b^T\Sig_y\b=1.
%\end{eqnarray}
On the other hand, if $\Sig$ is rank deficient, then choosing $\a$ and $\b$ as the upper and lower partitions of any vector in its null space will lead to full correlation, i.e., $\rho=1$.

In practice, the covariance matrices $\Sig_x$ and $\Sig_y$ and cross covariance matrix $\Sig_{xy}$ are usually unavailable. Instead, multiple independent observations $\x_i$ and $\y_i$ for $i=1\ldots N$ are measured and used to construct sample estimates of the (cross) covariance matrices:
\begin{eqnarray}
  \hat\Sig_x=\frac{1}{N}\X^T\X,\quad
  \hat\Sig_y=\frac{1}{N}\X^T\Y,\quad
  \hat\Sig_{xy}=\frac{1}{N}\X^T\Y,
\end{eqnarray}
where $\X=[\x_1\ldots \x_N]$ and $\Y=[\y_1\ldots \y_N]$. Then, these empirical matrices are used in the CCA formulation:
\begin{eqnarray}\label{emp-cca}
 \rho\(\hat\Sig_x,\hat\Sig_y,\hat\Sig_{xy}\)=\max_{\a\neq\0,\b\neq\0}\frac{\a^T\hat\Sig_{xy}\b}{\sqrt{\a^T\hat\Sig_x\a}\sqrt{\b^T\hat\Sig_y\b}}.
\end{eqnarray}
Clearly, if $N$ is sufficiently large then this sample approach performs well. However, in many applications, the number of samples $N$ is not sufficient. In fact, in the extreme case in which $N<n+m$ the sample covariance is rank deficient and $\rho=1$ independently of the data. The standard approach in such cases is to regularize the covariance matrices and solve $\rho\(\hat\Sig_x+\epsilon_y\I,\hat\Sig_y+\epsilon_x\I,\hat\Sig_{xy}\)$
where $\epsilon_x>0$ and  $\epsilon_x>0$ are small tuning ridge parameters \cite{vinod:76}.

CCA can be viewed as a unified framework for dimensionality
reduction in multivariate data analysis and generalizes other
existing methods. It is a generalization of PCA which seeks the
directions that maximize the variance of $\x$, and addresses the
directions corresponding to the correlation between $\x$ and $\y$. A
special case of CCA is PLS which maximizes the covariance of $\x$
and $\y$ (which is equivalent to choosing $\Sig_x=\Sig_y=\I$).
Similarly, MLR normalizes only one of the multivariates. In fact,
the regularized CCA mentioned above can be interpreted as a
combination of PLS and CCA \cite{borga-unified}.

\section{Sparse CCA}\label{sec_sparse}
We consider the problem of sparse CCA, i.e., finding a pair of linear combinations of $\x$ and $\y$ with prescribed cardinality which maximize the correlation. Mathematically, we define sparse CCA
as the solution to
\begin{eqnarray}\label{scca}
\left\{\begin{array}{ll}
  \max_{\a\neq\0,\b\neq\0} & \frac{\a^T\Sig_{xy}\b}{\sqrt{\a^T\Sig_x\a}\sqrt{\b^T\Sig_y\b}} \\
  \text{s.t.}
   & \|\a\|_0\leq k_a\\
   & \|\b\|_0\leq k_b.
 \end{array}\right.
\end{eqnarray}
Similarly, sparse PLS and sparse MLR are defined as special cases of (\ref{scca}) by choosing $\Sig_x=\I$ and/or $\Sig_y=\I$.
In general, all of these problems are difficult combinatorial problems. In small dimensions, they can be solved using a brute force search over all possible sparsity patterns and solving the associated subproblem via GEVD. Unfortunately, this approach is impractical for even moderate sizes of data sets due its exponentially increasing computational complexity. In fact, it is a
generalization of sparse PCA which has been proven NP-hard \cite{daspremont-2007}. Thus, suboptimal but efficient approaches are in order and will be discussed in the rest of this section.

\subsection{Existing solutions}
We now briefly review the different approaches to sparse CCA that appeared
in the last few years. Most of the methods are based on the well
known LASSO trick in which the difficult combinatorial cardinality
constraints are approximated through the convex L1 norm. This
approach has shown promising performance in the context of sparse
linear regression \cite{lasso}. Unfortunately, it is not sufficient
in the CCA formulation since the objective itself is not concave.
Thus additional approximations are required to transform the
problems into tractable form.

Sparse dimensionality reduction of rectangular matrices was
considered in \cite{aspremontgjl05} by combining the LASSO trick
with semidefinite relaxation. In our context, this is exactly sparse
PLS which is a special case of sparse CCA. Alternatively, CCA can be formulated as two constrained simultaneous regressions ($\x$ on $\y$, and $\y$ on $\x$). Thus, an appealing approach to sparse CCA is to use LASSO penalized regressions. Based on this idea,
\cite{hardoontr07} proposed to approximate the non convex constraints using the infinity norm. Similarly,
\cite{waaijenborg:2007,Parkhomenko:2007} proposed to use two nested iterative LASSO type regressions.

There are two main disadvantages to the LASSO based techniques.
First, there is no mathematical justification for their
approximations of the correlation objective. Second, there is no
direct control over sparsity. Their parameter tuning is difficult as
the relation between the L1 norm and the sparsity parameters is
highly nonlinear. The algorithms need to be run for each possible
value of the parameters and it is tedious to obtain the full
sparsity path.

An alternative approach for sparse CCA is sparse Bayes learning
\cite{FyfeL06,TanF01}. These methods are based on the probabilistic
interpretation of CCA, i.e., its formulation as an estimation
problem. It was shown that using different prior probabilistic
models, sparse solutions can be obtained. The main disadvantage of
this approach is again the lack of direct control on sparsity, and
the difficulty in obtaining its complete sparsity path.

Altogether, these works demonstrate the growing interest in deriving
efficient sparse CCA algorithms aimed at large data sets with simple
and intuitive parameter tuning.

\subsection{Greedy approach}
A standard approach to combinatorial problems is the forward
(backward) greedy solution which sequentially picks (or drops) the
variables at each stage one by one. The backward greedy approach to
CCA was proposed in \cite{thompson:84} but no specific algorithm was
derived or analyzed. In modern applications, the number of
dimensions may be much larger than the number of samples and
therefore we provide the details of the more natural forward
strategy. Nonetheless, the backward approach can be derived in a
straightforward manner. In addition, we derive an efficient
approximation to the subproblems at each stage which significantly
reduces the computational complexity. A similar approach in the
context of PCA can be found in \cite{daspremont-2007}.

Our goal is to find the two sparsity patterns, i.e., two sets of indices  $I$ and $J$ corresponding to the indices of the chosen variables in $\x$ and $\y$, respectively. The greedy algorithm chooses the first elements in both sets as the solution to
\begin{eqnarray}
 \max_{i,j}\frac{\Sig_{xy}^{i,j}}{\sqrt{\Sig_x^{ii}}\sqrt{\Sig_y^{jj}}}.
\end{eqnarray}
Thus, $I=\{i\}$ and $J=\{j\}$. Next, the algorithm sequentially
examines all the remaining indices and computes
\begin{eqnarray}\label{addi}
 \max_{i\notin I}\rho\(\Sig_x^{I\cup i,I\cup i},\Sig_y^{J,J},\Sig_{xy}^{I\cup i,J}\)
\end{eqnarray}
and
\begin{eqnarray}\label{addj}
 \max_{j\notin J}\rho\(\Sig_x^{I,I},\Sig_y^{J\cup j,J\cup j},\Sig_{xy}^{I,J\cup j}\).
\end{eqnarray}
Depending on whether (\ref{addi}) is greater or less than (\ref{addj}), we add index $i$ or $j$ to $I$ or
$J$, respectively. We emphasize that at each stage, only one
index is added either to $I$ or $J$. Once one of the set reaches its
required size $k_a$ or $k_b$, the algorithm continues to add indices
only to the other set and terminates when this set reaches its
required size as well. It outputs the full sparsity path, and returns $k_a-k_b-1$ pairs of vectors associated with the sparsity patterns in each of the stages.

The computational complexity of the algorithm is polynomial in the dimensions of the problem. At each of its $i=1,\cdots,k_a+k_b-1$ stages, the algorithm computes $n+m-i$ CCA solutions as expressed in (\ref{addi}) and (\ref{addj}) in order to select the patterns for the next stage. It is therefore reasonable for small problems, but is still impractical for many applications. Instead, we now propose an alternative approach that computes only one CCA per stage and reduces the complexity significantly.

An approximate greedy solution can be easily obtained by approximating (\ref{addi}) and (\ref{addj}) instead of solving them exactly. Consider for example (\ref{addi}) where index $i$ is added to the set $I$. Let $\a^{I,J}$ and $\b^{I,J}$ denote the optimal solution to $\rho\(\Sig_x^{I,I},\Sig_y^{J,J},\Sig_{xy}^{I,J}\)$. In order to evaluate (\ref{addi}) we need to recalculate both $\a^{I\cup i,J}$ and $\b^{I\cup i,J}$ for each $i\notin I$. However, the previous $\b^{I,J}$ is of the same dimension and still feasible. Thus, we can optimize only with respect to $\a^{I\cup i,J}$ (whose dimension has increased). This approach provides the following bounds:

\begin{lemma}
Let $\a^{I,J}$ and $\b^{I,J}$ be the optimal solution to $\rho\(\Sig_x^{I,I},\Sig_y^{J,J},\Sig_{xy}^{I,J}\)$ in (\ref{cca}). Then,
\begin{eqnarray}\label{lemma}
 \rho^2\(\Sig_x^{I\cup i,I\cup i},\Sig_y^{J,J},\Sig_{xy}^{I\cup i,J}\)
  \!\!\!&\!\!\!\geq\!\!\!&\!\!\!{ \rho^2\(\Sig_x^{I,I},\Sig_y^{J,J},\Sig_{xy}^{I,J}\)
 +\delta^{I,J}_i}\nonumber\\
 \rho^2\(\Sig_x^{I,I},\Sig_y^{J\cup j,J\cup j},\Sig_{xy}^{I,J\cup j}\)
  \!\!\!&\!\!\!\geq\!\!\!&\!\!\!{ \rho^2\(\Sig_x^{I,I},\Sig_y^{J,J},\Sig_{xy}^{I,J}\)
 +\gamma^{I,J}_j},\!\!\!\!\!\!\!\!\nonumber\\
\end{eqnarray}
where
\begin{eqnarray}\label{deltas1}
  \delta^{I,J}_i\!=\!\frac{\(\[\b^{I,J}\]^T\!\!\[\Sig_{xy}^{I,J}\]^T\!\!\[\Sig_x^{I,I}\]^{-\!1}\!\Sig_x^{I,i}\!-\!\!\[\b^{I,J}\]^T\!\!\[\Sig_{xy}^{i,J}\]^T\)^2}
 {\Sig_x^{i,i}-\[\Sig_x^{I,i}\]^T\[\Sig_x^{I,I}\]^{-1}\Sig_x^{I,i}}\nonumber
\end{eqnarray}
\begin{eqnarray}\label{deltas2}
 \gamma^{I,J}_j\!=\!\frac{\(\[\a^{I,J}\]^T\!\!\Sig_{xy}^{I,J}\[\Sig_y^{J,J}\]^{-\!1}\!\Sig_y^{J,j}-\[\a^{I,J}\]^T\!\!\Sig_{xy}^{I,j}\)^2}
 {\Sig_y^{j,j}-\[\Sig_y^{J,j}\]^T\[\Sig_y^{J,J}\]^{-1}\Sig_y^{J,j}},
\end{eqnarray}
and we assume that all the involved matrix inversions are nonsingular.
\end{lemma}
Before proving the lemma, we note that it provides lower bounds on the increase in cross correlation due to including an additional element in $\x$ or $\y$ without the need of solving a full GEVD. Thus, we propose the following approximate greedy approach. For each sparsity pattern $\{I,J\}$, one CCA is computed via GEVD in order to obtain $\a^{I,J}$ and $\b^{I,J}$. Then, the next sparsity pattern is obtained by adding the element that maximizes $\delta^{I,J}_i$ or $\gamma^{I,J}_j$ among $i\notin I$ and $j\notin J$.

\begin{proof}
We begin by rewriting (\ref{cca}) as a quadratic maximization with constraints. Thus, we define $\a^{I,J}$ and $\b^{I,J}$ as the solution to
\begin{eqnarray}\label{app1}
 \rho\(\Sig_x^{I,I},\Sig_y^{J,J},\Sig_{xy}^{I,J}\)=\left\{\begin{array}{ll}
  \max_{\a,\b} & \a^T\Sig_{xy}^{I,J}\b \\
  \text{s.t.} & \a^T\Sig_x^{I,I}\a=1 \\
   & \b^T\Sig_y^{J,J}\b=1.
 \end{array}\right.
\end{eqnarray}
Now, consider the problem when we add variable $j$ to the set $J$:
\begin{eqnarray}\label{app2}
 \rho\(\Sig_x^{I,I},\Sig_y^{J\cup j,J\cup j},\Sig_{xy}^{I,J\cup j}\)=\left\{\!\!\!\begin{array}{ll}
  \max_{\a,\b} & \a^T\Sig_{xy}^{I,J\cup j}\b \\
  \text{s.t.} & \a^T\Sig_x^{I,I}\a=1 \\
   & \b^T\Sig_y^{J\cup j,J\cup j}\b=1.
 \end{array}\right.\!\!\!\!\!\!\!\!\!\!\!\!\!\!\!\!\!\!\!\nonumber\\
\end{eqnarray}
Clearly, the vector $\a=\a^{I,J}$ is still feasible (though not necessarily optimal) and yields a lower bound:
\begin{eqnarray}
\rho\(\Sig_x^{I,I},\Sig_y^{J\cup j,J\cup j},\Sig_{xy}^{I,J\cup j}\)\geq\left\{\!\!\!\begin{array}{ll}
  \max_{\b} & \[\a^{I,J}\]^T\Sig_{xy}^{I,J\cup j}\b \\
  \text{s.t.} &  \b^T\Sig_y^{J\cup j,J\cup j}\b=1.
 \end{array}\right.\!\!\!\!\!\!\!\!\!\!\!\!\!\!\!\!\!\!\!\nonumber\\
\end{eqnarray}
Changing variables $\overline{\b}=\[\Sig_y^{J\cup j,J\cup j}\]^{\frac{1}{2}}\b$ results in:
\begin{eqnarray}
\rho\(\Sig_x^{I,I},\Sig_y^{J\cup j,J\cup j},\Sig_{xy}^{I,J\cup j}\)\geq\left\{\begin{array}{ll}
  \max_{\overline{\b}} & \c^T\overline{\b} \\
  \text{s.t.} &  \|\overline{\b}\|_2=1,
 \end{array}\right.
\end{eqnarray}
where
\begin{eqnarray}
  \c=\[\Sig_y^{J\cup j,J\cup j}\]^{-\frac{1}{2}}\[\Sig_{xy}^{I,J\cup j}\]^T\a^{I,J}.
\end{eqnarray}
Using the Cauchy Schwartz inequality
\begin{eqnarray}
  \c^T\overline{\b}\leq\|\c\|_2\|\overline{\b}\|_2,
\end{eqnarray}
we obtain
\begin{eqnarray}
&&\rho\(\Sig_x^{I,I},\Sig_y^{J\cup j,J\cup j},\Sig_{xy}^{I,J\cup j}\)\nonumber\\
&&\qquad\qquad\geq\left\|\[\Sig_y^{J\cup j,J\cup j}\]^{-\frac{1}{2}}\[\Sig_{xy}^{I,J\cup j}\]^T\a^{I,J}\right\|_2.
\end{eqnarray}
Therefore,
\begin{eqnarray}
&&\rho^2\(\Sig_x^{I,I},\Sig_y^{J\cup j,J\cup j},\Sig_{xy}^{I,J\cup j}\)\geq\[\a^{I,J}\]^T\[\begin{array}{cc}
                           \Sig_{xy}^{I,J} & \Sig_{xy}^{I,j}
                         \end{array}\]\nonumber\\
&&\qquad\times\[\begin{array}{cc}
             \Sig_y^{J,J} & \Sig_y^{J,j} \\
             \Sig_y^{J,J} & \Sig_y^{J,j}
           \end{array}
   \]^{-1}\[\begin{array}{c}
                           \[\Sig_{xy}^{I,J}\]^T \\ \[\Sig_{xy}^{I,j}\]
                         \end{array}\]^T\a^{I,J}.
\end{eqnarray}
Finally, (\ref{lemma}) is obtained by using the inversion formula for partitioned matrices and simplifying the terms.
\end{proof}

\section{Numerical results}\label{sec_num}
\begin{figure}
\center
\includegraphics[width=0.35\textwidth]{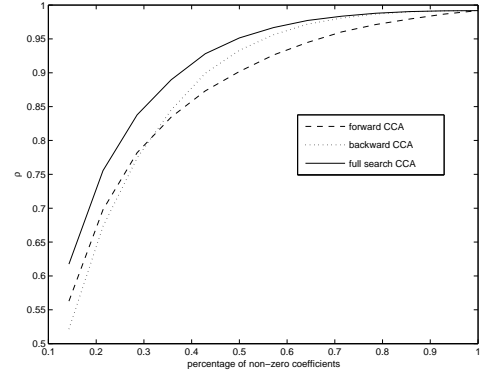}
\caption{Correlation vs. Sparsity, $n=m=7$. \label{figsparsity}}
\end{figure}

We now provide a few numerical examples illustrating the behavior of
the greedy sparse CCA methods. In all of the simulations below, we implement the greedy methods using the bounds in Lemma 1. In the first experiment we evaluate
the validity of the approximate greedy approach. In particular, we choose $n=m=7$, and generate $200$
independent random realizations of the joint covariance matrix using the Wishart distribution with $7+7$ degrees of freedom.
For each realization, we run the approximate greedy
forward and backward algorithms and calculate the full sparsity
path. For comparison, we also compute the optimal sparse solutions using an exhaustive search. The results are presented in Fig. \ref{figsparsity} where the
average correlation is plotted as a function of the number of variables (or
non-zero coefficients). The greedy methods capture a significant portion of the possible correlation.
As expected, the forward greedy approach outperforms the backward
method when high sparsity is critical. On the other hand, the
backward method is preferable if large values of correlation are
required.

\begin{figure}
\center
\includegraphics[width=0.35\textwidth]{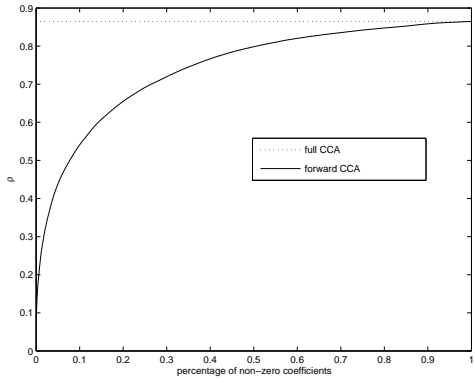}
\caption{Correlation vs. Sparsity, $n=m=1000$. \label{figsparsity2}}
\end{figure}

In the second experiment we demonstrate the performance of the approximate forward greedy approach in a large scale problem. We present results for a representative (randomly generated) covariance matrix of sizes $n=m=1000$. Fig. \ref{figsparsity2} shows the full sparsity path of the greedy method. It is easy to see that about 90 percent of the CCA correlation
value can be captured using only half of the variables.
Furthermore, if we choose to capture only 80 percent of the full
correlation, then about a quarter of the variables are sufficient.

\begin{figure}
\center
\includegraphics[width=0.35\textwidth]{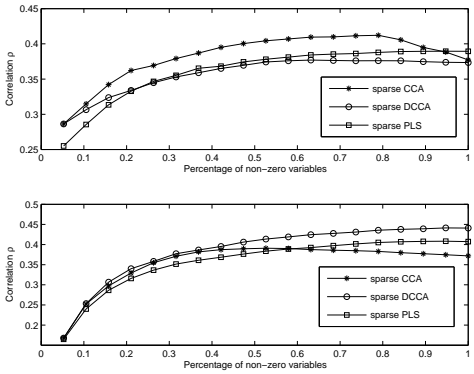}
\caption{Sparsity as a regularized CCA method, $n=10$, $m=10$ and $N=20$.
\label{figreg}}
\end{figure}

In the third set of simulations, we examine the use of sparse
CCA algorithms as regularization methods when the number of samples
is not sufficient to estimate the covariance matrix efficiently. For simplicity, we restrict
our attention to CCA and PLS (which can be interpreted as an extreme case of ridge CCA). In addition, we show results for an
alternative method in which the sample covariance $\hat\Sig_x$ and
$\hat\Sig_y$ are approximated as diagonal matrices with the sample
variances (which are easier to estimate). We refer to this method as
Diagonal CCA (DCCA). In order to assess the regularization
properties of CCA we used the following procedure. We randomly
generate a single ``true'' covariance matrix $\Sig$ and use it
throughout all the simulations. Then, we generate $N$ random
Gaussian samples of $\x$ and $\y$ and estimate $\hat\Sig$. We apply
the three approximate greedy sparse algorithms, CCA, PLS and DCCA, using the
sample covariance and obtain the estimates $\hat\a$ and $\hat\b$.
Finally, our performance measure is the ``true'' correlation value
associated with the estimated weights which is defined as:
\begin{eqnarray}\label{hatrho}
\frac{\hat\a^T\Sig_{xy}\hat\b}{\sqrt{\hat\a^T\Sig_x\hat\a}\sqrt{\hat\b^T\Sig_y\hat\b}}.
\end{eqnarray}
We then repeat the above procedure (using the same ``true''
covariance matrix) 500 times and present the average value of
(\ref{hatrho}) over these Monte Carlo trials. Fig. \ref{figreg} provides
these averages as a function of parsimony for two representative
realizations of the ``true'' covariance matrix. Examining the
curves reveals that variable selection is indeed a promising
regularization strategy. The average correlation increases with the
number of variables until it reaches a peak. After this peak, the
number of samples are not sufficient to estimate the full covariance
and it is better to reduce the number of variables through sparsity. DCCA can also be slightly improved by using fewer
variables, and it seems that PLS performs best with no subset
selection.

\section{Discussion}\label{sec_conc}
We considered the problem of sparse CCA and discussed its
implementation aspects and statistical properties. In particular, we
derived direct greedy methods which are specifically designed to
cope with large data sets. Similar to state of the art sparse
regression methods, e.g., Least Angle Regression (LARS) \cite{lars},
the algorithms allow for direct control over the sparsity and
provide the full sparsity path in a single run. We have
demonstrated their performance advantage through numerical
simulations.

There are a few interesting directions for future research in sparse
CCA. First, we have only addressed the first order sparse canonical
components. In many applications, analysis of higher order canonical components is preferable. Numerically, this extension can be implemented by subtracting the first components and
rerunning the algorithms. However, there remain interesting
theoretical questions regarding the relations between the
sparsity patterns of the different components.
Second, while here we considered the case of a pair of multivariates, it
is possible to generalize the setting and address multivariate correlations
between more than two data sets.

\end{document}